\pdfoutput=1
\documentclass[a4paper,14pt]{extreport}
\usepackage[utf8]{inputenc}
\usepackage[english,russian]{babel}
\usepackage[dvips]{graphicx}
\usepackage{geometry}
\usepackage{epsfig} 
\usepackage{fancyhdr} 
\usepackage{amsmath}
\usepackage{theorem}
\usepackage{subfigure} 
\usepackage{fxspace} 
\usepackage{listings} 
\usepackage{amssymb} 
\usepackage{makeidx}
\usepackage{hyperref}
\usepackage[usenames]{color}
\usepackage{colortbl}
\usepackage{multirow}
\usepackage{setspace}
\usepackage[toc,page]{appendix}
\usepackage{lipsum}

\graphicspath{{img/}}

\newcommand\tab{\hspace{6mm}}

\newcommand{\articlenameru}{Статическая инструментация байткода виртуальной машины Dalvik}
\newcommand{\articlename}{Static Dalvik VM bytecode instrumentation}

\newenvironment{localsize}[1]
{%
  \clearpage
  \let\orignewcommand\newcommand
  \let\newcommand\renewcommand
  \makeatletter
  \input{bk#1.clo}%
  \makeatother
  \let\newcommand\orignewcommand
}
{%
  \clearpage
}

\renewcommand{\chaptermark}[1]%
	{\markboth{\chaptername~\thechapter~--~#1}{}}

\renewcommand{\sectionmark}[1]%
	{\markright{\thesection\ #1}}

\begin{document}
\let\ref\autoref

\begin{localsize}{12}
\begin{titlepage}
\begin{table}
    \centering
    \begin{tabular}{rcl}
    Автономная некоммерческая &
    \multirow{4}{*}{\includegraphics[width=40mm]{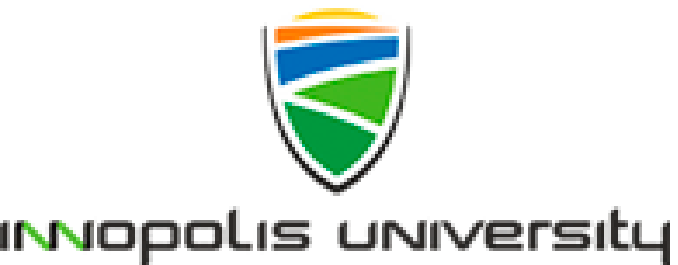}}
    & Autonomous noncommercial \\
    организация высшего  & & organization of higher \\
    образования & & education \\
    «Университет Иннополис»  &     
     & «Innopolis University» \\
    \hline
    \hline
    \end{tabular}
    \label{tab:my_label}
\end{table}
\vline
\vspace{5mm}

\begin{center}
\textbf{
ВЫПУСКНАЯ КВАЛИФИКАЦИОННАЯ РАБОТА  \\
ПО НАПРАВЛЕНИЮ ПОДГОТОВКИ \\
09.03.01 --- «ИНФОРМАТИКА И ВЫЧИСЛИТЕЛЬНАЯ ТЕХНИКА»}
\vspace{5mm}

\textbf{GRADUATE THESIS    \\
MAJOR: "COMPUTER SCIENCE"}
\end{center}
\vspace{20mm}

    \begin{tabular}{ll
|>{\columncolor[gray]{.8}}l|}
\cline{3-3}
\textbf{Тема:} &
    \makebox[0.5mm] &
    \makebox[135mm][l]{\articlenameru}    \\
    &&\\
    && \\
    &&  \\
\cline{3-3}
    \end{tabular}
\vspace{5mm}

    \begin{tabular}{ll
|>{\columncolor[gray]{.8}}l|}
\cline{3-3}
\textbf{Topic:} &
     &
    \makebox[135mm][l]{\articlename}    \\
    &&\\
    && \\
    &&  \\
\cline{3-3}
    \end{tabular}
\vspace{5mm}

    \begin{tabular}{ll
|>{\columncolor[gray]{.8}}l|l
|>{\columncolor[gray]{.8}}l|}
\cline{3-3} \cline{5-5}
Работу выполнил / &
    \makebox[0.5mm] &
    \makebox[64mm][l]{Минибаев Евгений Маратович /}  &
    &       \\
Thesis is executed by  &
    \makebox[0.5mm] &
    \makebox[64mm][l]{Eugene Minibaev Maratovich}  &
    &
    \makebox[40mm]{\textcolor[gray]{.6}{подпись / signature}}     \\
\cline{3-3} \cline{5-5}  \\
    \end{tabular}
\vspace{5mm}

    \begin{tabular}{ll
|>{\columncolor[gray]{.8}}l|l
|>{\columncolor[gray]{.8}}l|}
\cline{3-3} \cline{5-5}
Научный руководитель / &
     &
    \makebox[57mm][l]{Нестор Катано Коллазос /}  &
    &       \\
Thesis supervisor  &
     &
    \makebox[57mm][l]{N{\'e}stor Cata{\~n}o Collazos}  &
    &
    \makebox[40mm]{\textcolor[gray]{.6}{подпись / signature}}     \\
\cline{3-3} \cline{5-5}  \\
    \end{tabular}
\vspace{\fill}

\begin{center}
Иннополис, Innopolis, 2017
\end{center}
\end{titlepage}

\end{localsize}

\selectlanguage{english}
\thispagestyle{empty}
~
\newpage

\setcounter{page}{3}

\tableofcontents
\addcontentsline{toc}{chapter}{List of figures}
\listoffigures

 \chapter{Abstract}
\label{ChapAbstract}




\tab This work proposes a novel approach to restricting the access for blacklisted Android system API calls. Main feature of the suggested method introduced in this paper is that it requires only rootless or (user-mode) access to the system unlike previous works. For that reason this method is valuable for end-users due to the possibility of project distribution via Play Market and it does not require any phone system modifications and/or updates.
This paper explains the required background of Android OS necessary for understanding and describes the method for modification Android application. In this paper the proof-of-concept implementation. That is able to block the application's IMEI requests is introduced. Also this paper lists unsuccessful methods that tried to provide the user security. Obviously with those restrictions application may lack some of features that can only be granted in unsecured environment.


 \chapter{Introduction}
\label{ChapIntro}



\begin{figure}[!h]
  \centering
    \includegraphics[width=0.8\textwidth]{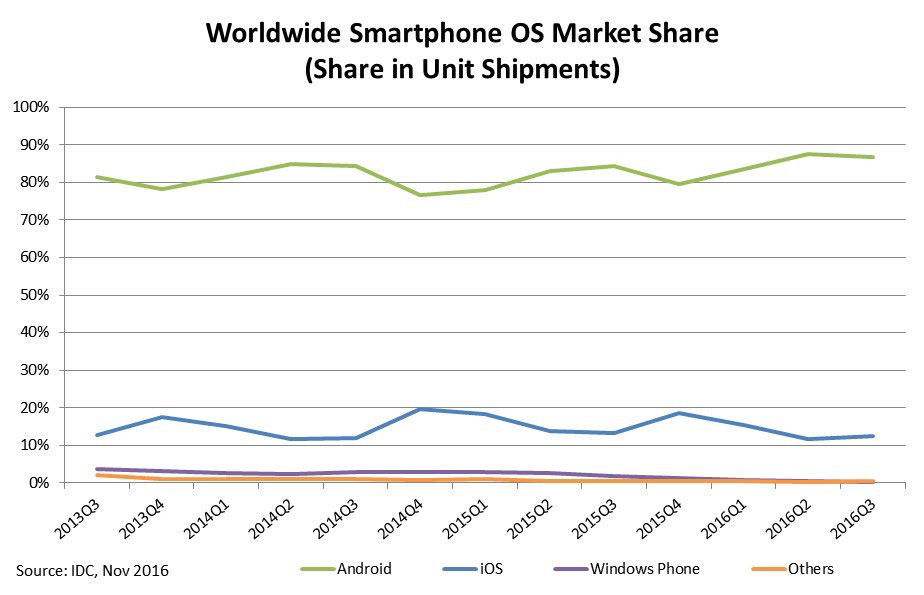}
  \caption{Mobile OS market share}
  \label{os_market_share}
\end{figure}

\tab For the past 8 years Google, Inc and Open Handset Alliance confidently remain undisputed leaders in the market of mobile operating systems. In the middle of the 2016 their well-known product named Android has almost 90\% of mobile OS market share~\ref{os_market_share} sold to the end users as for 2015 there were nearly 1\.4 billion active devices working under this operating system~\cite{14devices}. Therefore it is essential to be ensured in the quality of the security and user experience of that platform.

One of the important features of this platform is that anyone can create and distribute applications with the instruments and libraries provided by the Android ecosystem. Developers can implement their apps with either Java (SDK) or C/C++ (NDK) development kits and then distribute it via the Google Play.


Obviously most of the applications will be useless if they cannot be able to work with the personal data and Android platform isn't an exception. As long as developers make mistakes that implies the problem of improper handling of that important data. Some studies show that it often takes the place in the real world applications~\cite{chin2011analyzing} \cite{enck2014taintdroid}.

Another aspect of that issue is that developers usually require more access than their application really needs. Despite the attempts to connect semantic purpose of the application and the requested permissions\cite{slavin2016toward}, the reality shows that meta information of the application usually is not enough~\cite{benats2011primandroid}.


From user's perspective it may seem that some instant messenger applications, which requires for their work the camera and the SMS reading permissions, don't seem legitimate. But still those can be used in the proper way. For example the camera permission may improve the quality of the photo and SMS reading permission may be used for the account authentication. Therefore it seems impossible (or almost impossible) to distinguish the malicious and benign permissions.


User decision can be taken into account as a possible solution for that problem. But the problem is that Android doesn't have such possibilities. The solution proposed in this paper presents a technique that gives a user the possibility to deny selected permissions of the application.


The chapter of \texttt{Background theory} presents essential technical information required for complete understanding of the solution. The following chapter of \texttt{Results} presents the output of the research and describes the implemented proof of concept. As long as the area of the Android security is not new, the last chapter of \texttt{Related Works} discusses the similar works that have been done on this subject or somehow bounded to this issue. The possible consequences for research on this subject is in the last chapter \texttt{Future Works}.

 \chapter{Background theory}
\label{ChapBackgroundtheory}
\section{Android Application structure}

\begin{figure}[!h]
  \centering
    \includegraphics[width=0.8\textwidth]{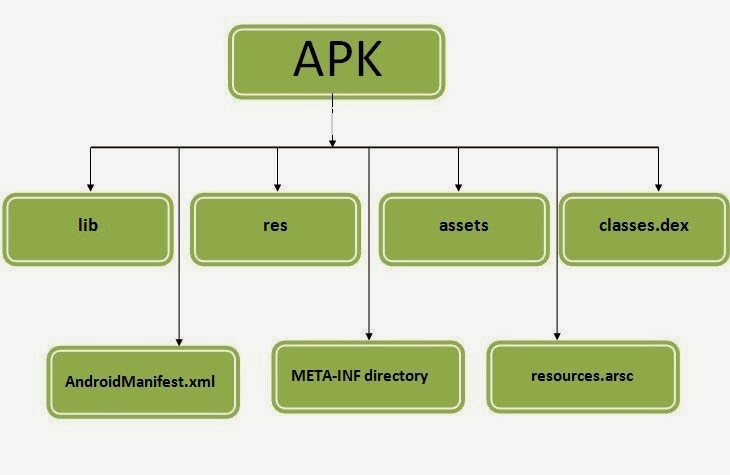}
  \caption{Android APK structure}
  \label{android_apk}
\end{figure}

\tab Every Android application is a APK zip archive similar to Java jar files. You can see its contents at the~\ref{android_apk}. This files are distributed via official Google Play application that is installed with the OS.

The main code of the application is placed to the \textit{classes.dex} file. This is a special bytecode file format that is supposed to run on the Dalvik Virtual Machine. It consists of the compiled \textit{.class} files that are built on the Java code and are represented in more a optimized way. It is also possible to write a native code in C/C++ languages with the Android NDK. Finally the code is compiled in the Linux shared libraries (.so files) and placed to the \textit{lib} directory. The last one is assessed via Java Native Interface (JNI) same as in usual Java.

The \textit{AndroidManifest.xml} contains meta information of the packages in binary XML format. This, for example, includes the main Activity class and permissions information. The same way as in the jar files the \textit{META-INF} contains signatures and hashes of the files in the archive.

The \textit{res} and \textit{assets} directory contains resources data and \textit{resources.arsc}, and the meta information of both.

\section{Resources}

The developers of Android application can use resources for extraction of the application data and to ease the development process. Those are written in the XML format and may contain some constants, graphics, application layouts, etc. All the access in the code is performed via fixed class \textbf{R} similar to Java static field access in form \textbf{R.$<$type$>$.$<$name$>$}, e.g: \textbf{R.string.my-string}. During the compilation the compilation types created as nested classes of \textbf{R}, and the access is exactly as to the static field.

\section{Android Components}

\tab Android application is built on one or more components. There are four types of such building blocks~\cite{AndroidComponents}:

\begin{itemize}
    \item \textit{Activities} is the entry point for the main interacting interface with the user. It represents a single screen with a user interface.
    \item \textit{Services} is a general-purpose entry point for keeping an app running in the background for all kinds of reasons. It is a component that runs in the background to perform long-running operations or to perform work for remote processes. A service does not provide a user interface.
    \item \textit{Broadcast Receivers} is a component that enables the system to deliver events to the app outside of a regular user flow, allowing the app to respond to system-wide broadcast announcements.
    \item \textit{Content Providers} manages a shared set of app data that applications can store in the file system, in a SQLite database, on the web, or on any other persistent storage location that app can access. 
\end{itemize}

\section{Messaging system}

\tab Android OS provides a quite complicated communication system between components. It is called an \textbf{Intent}. It is simply a message that contain an address and data for transmission. These message objects can be used both for inner- and inter-application communication. Besides those variants such messages can be sent by operating system.

\section{Activity lifecycle}

\begin{figure}[!htb]
  \centering
    \includegraphics[width=0.7\textwidth]{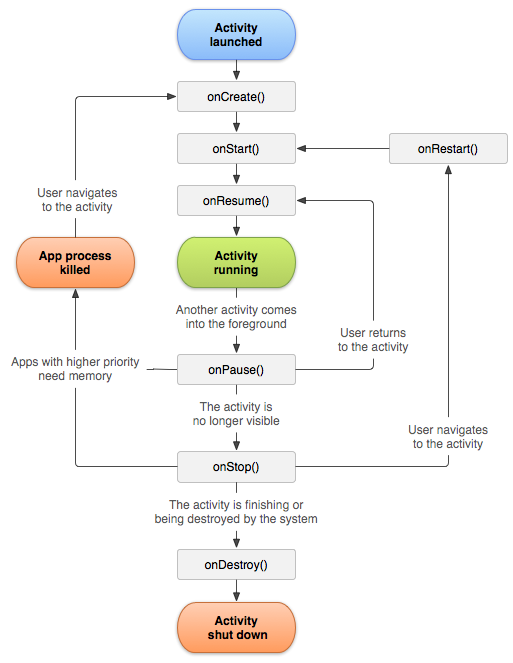}
  \caption{Activity Lifecycle}
  \label{activity_lifecycle}
\end{figure}

\tab In the application there may be many entry points and all of them are independent. The overall developing concept is reactive that means the application should react on the events via callbacks. Some of them are \textbf{onCreate}, \textbf{onPause}, \textbf{onDestroy} and the rest is presented at~\ref{activity_lifecycle}. In other words the Activity flow represents a state machine where the transitions are events and the states are callbacks provided by developers and they are executed on the event. From this flow it is also important to mention that the application is notified when they're going to suspend and resume unlike in other popular OSes.

The application launch can be performed via corresponding system call with some predefined intent messages. This also includes an example of the user interaction with launching apps via menu application.

\section{Manifest Permissions}

\tab For performing system calls the Android OS provides the permission system. During the installation a user sees the requested permissions and can decide whether to accept them and install the application, or to deny those requests. This list of such requirements is placed to the \textit{AndroidManifest.xml} file.

There is a big list of system permissions. In addition developers can also define their own. All of them are divided into four protection levels \cite{AndroidProtectionLevel}:


\begin{itemize}
    \item \textit{Normal} permissions cover areas where app needs to access data or resources outside the app's sandbox, but where there's very little risk to the user's privacy or the operation of other apps. The system automatically grants the permission of that level to the app.
    \item \textit{Dangerous} permissions cover areas where the app wants data or resources that involve the user's private information, or could potentially affect the user's stored data or the operation of other apps. Those ones are granted after user confirmation.
    \item \textit{Signature} permission that the system grants only if the requesting application is signed with the same certificate as the application that declared the permission.
    \item \textit{Signature or system} permission that the system grants only to applications that are in the Android system image or that are signed with the same certificate as the application that declared the permission.
\end{itemize}


One of the problems with the permissions is that they should be either applied all at once, or nothing at all. So a user cannot accept only some pieces of these requests. The developers of the Android OS also understand this problem. But because they are very limited in the modification of the architecture they cannot change a lot.

Despite the beginning with the \textbf{Android~6.0} developers can ask for some required permissions in runtime~\cite{android0doc0runtime0perm}. But it is still not really practical for two reasons. Firstly, due to platform version compatibility developers still may write the code in the old way and ask all required permissions during the installation. The second problem is that only nearly a third~\cite{android0doc0dashboards} of all existing devices is able to use runtime permissions.

\section{Android system architecture}

\tab Android operating system is a number of components built on the modified Linux kernel and most of the features of Linux derived to that mobile OS. Also  Android developers was very mindful about the architectural design of the platform in general and so this platform is one of the most secure from the user's perspective~\cite{faruki2015android}.

\begin{figure}[!h]
  \centering
    \includegraphics[width=0.55\textwidth]{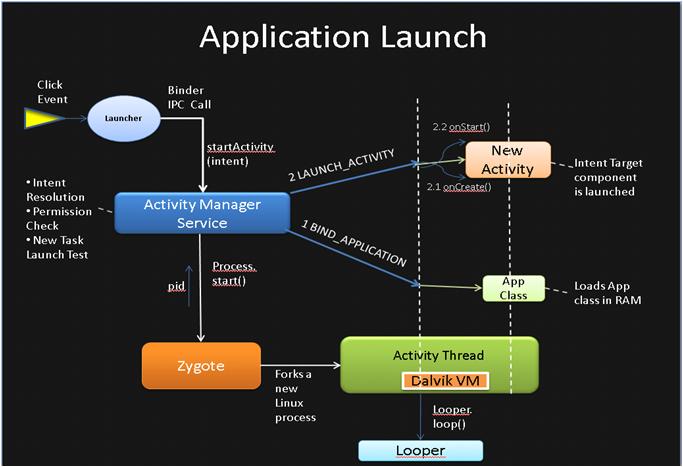}
  \caption{Application launch}
  \label{app_launch}
\end{figure}

All applications are separated from each other. This is enforced by DAC (Discretionary Access Control) from Linux kernel by assigning each app unique user and private directory. Besides every application is running in its own copy of Dalvik VM (or DVM)~\cite{ehringer2010dalvik}. DVM is an alternative to Java virtual machine for the dex bytecode. The process that copies such VMs is called Zygote. Its job is to perform Linux \textbf{fork()} system call and pass the control to the app. An overall controlling is performed via ActivityManager. You can find more complete application launch system flow at the~\ref{app_launch}.

\section{Dex files}

\begin{figure}[!htb]
  \centering
    \includegraphics[width=0.55\textwidth]{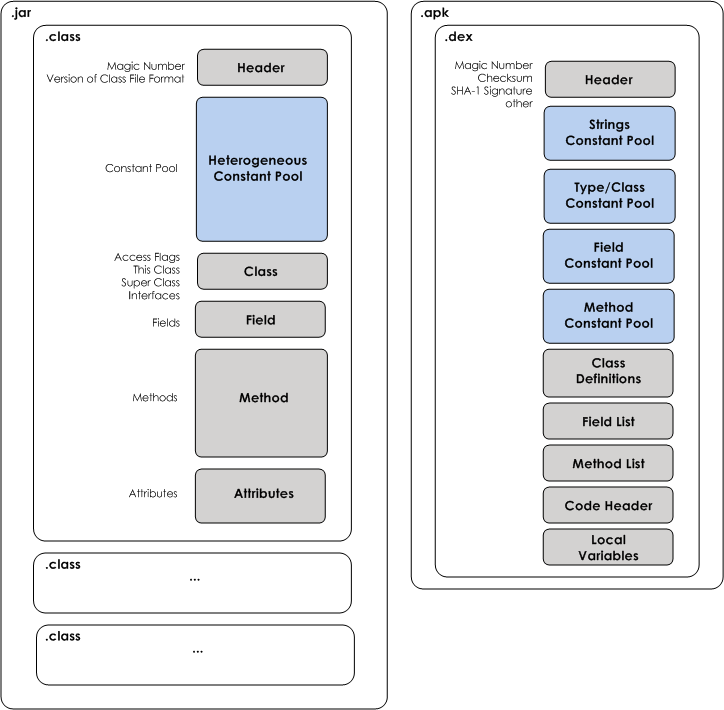}
  \caption{JAR and APK comparison}
  \label{jar_apk_compare}
\end{figure}

The \textit{class.dex} files represents the code of the application in the Dalvik bytecode format. It can be compared to \textit{\*.class} Java bytecode files. The dex file is actually excluded from many \textit{\*.class} files (\ref{jar_apk_compare}) in a more optimized way, because strings don't need to be duplicated in one file.

The dex file consists of a header, the following table of strings, class definitions, and then indexes. The instructions of Dalvik bytecode are based on these indexes. So if we want to search for some function \textbf{MyClass.func}. We need to search for a class definition of \textbf{MyClass} in the method list. And then based on an index in that list and in the method table to find the method that maps to string ``\textbf{func}''. The code example that can traverse these tables and print method signature of the class you can find at~\ref{appendix:class_methods}. The real instruction of the bytecode work with the index in the method list.

\begin{table}[!hbt]
  \centering\begin{tabular}{|c|c|c|}
    \hline
    opcode          & prefix & instruction size\\
    \hline
    invoke-virtual          & 0x6e & 6 bytes\\
    invoke-super            & 0x6f & 6 bytes\\
    invoke-direct           & 0x70 & 6 bytes\\
    invoke-static           & 0x71 & 6 bytes\\
    invoke-interface        & 0x72 & 6 bytes\\
    invoke-virtual/range    & 0x74 & 6 bytes\\
    invoke-super/range      & 0x75 & 6 bytes\\
    invoke-direct/range     & 0x76 & 6 bytes\\
    invoke-static/range     & 0x77 & 6 bytes\\
    \hline
  \end{tabular}
  \caption{invoke bytecode listing}
  \label{invokes}
\end{table}

The function call is performed via invoke-kind opcode family. It is important to mention that all invoke-kind bytecode instructions have the same size:~\ref{invokes}. So after patching method call with other type of invoke code doesn't need to be re-aligned and resulted bytes remains the same. The patching in proof-of-concept is implemented in the such way.



 \chapter{Results}
\label{ChapResults}




The main research was held in the area of giving the user a possibility of limiting the application permissions requests. One of the interesting points of Android is that it doesn't forbid access to the installed application code as well as listing installed application. Therefore it is possible to modify the code of the specified application. After researching the operating system main possible scenarios have been discovered:

\begin{itemize}
  \item Steal the Zygote work: perform fork and then app loading to the new process
  \item The process substitution: loading of the application and substitution of the current process with the new one
  \item Application repacking and reinstalling
\end{itemize}

All of these methods include the APK loading, searching for the dangerous calls and patching them with the secure versions. The following sections describe the results of each method.

The other issue that occurred is that because the code in fact is a memory mapped structure comprising the whole code of the application, it is not possible to modify the code of a class dynamically. In fact, because of the second constraint, it would be hard to modify a single class anyway, without modifying the code of the other classes. Therefore the modification of the bytecode cannot be performed on the already launched code.

\section{Steal the Zygote work}

Because all application start with the different users and therefore run in the unaccessible memory we cannot directly modify the running application data.
So the essential way of launching tampered application code is a becoming a Zygote itself. As long as the work of the Zygote is quite simple it shouldn't be difficult.

Unfortunately \textbf{fork()} is not recommended for usage for user processes because it is internal API. The consequent issue with the application loading is described in the following section.

\section{The process substitution}

If we cannot create new processes for the tampered application we can still attempt to load the code and to pass all the callbacks to the new instance of Activity.

In practice this is also a hard way because it should initialize the private members of Activity instance and the thread information. Also there is a good probability of collision resource classes during the loading.

\section{Application repacking}

As long as more ``dynamic'' methods don't work or are too complicated the last (fortunately successful) attempt was the repacking of the application and its following re-installation.

The flow of this idea is pretty simple. Firstly, it should find the requested apk file of the installed package. The second stage is analyzing the \textit{classes.dex} file and generating required stub methods. If we want to forbid \textbf{TelephonyManager.getDeviceId} that gets the IMEI of the phone we need to generate the static method with the same return parameter, for this example it should return \textbf{java.lang.String}, it's not really necessary to handle function parameters in general if the finite aim is to forbid calls.

The proof-of-concept code can be found at \textit{https://github.com/l4l/bs-thesis}. The code architectural structure is the simplest possible. The ApkLoader is the main package that is able to patch, repack and sign the packages. Bytecode patching is performed via MethodBuffer class. This class works with the given buffer of dex file and performs patching method calls.

Obviously this work is far from ideal, for example simple obfuscation techniques don't allow to perform such kind of modifications.

\section{Implementation dependencies}

The Android platform has a nice tool for merging bunch of class files to the dex file. Because it also has an implementation of bytecode generation that was enough for implementing the patching and repacking. The bytecode generation example of stub class and empty method is provided in~\ref{appendix:stub}.  

The other additional library is zipsigner, so it is possible to perform the signing.

 \selectlanguage{english}

\chapter{Related Work}
\label{ChapRelated}


The area of the research was pretty wide so it includes the many areas of the topic. Beginning with the basic communication analysis vulnerabilities~\cite{chin2011analyzing} and ending with enforcement security via kernel modification~\cite{nauman2010apex}\cite{balavz2015android}.

The other way is a dynamic approach that is presented in Aurasium paper~\cite{xu2012aurasium}. They propose enforcing policies via system calls hooking. Their work is based on the insertion of the arbitrary native library to the APK file so the framework is able to enforce the policies. The enforcement is based on the rewriting the main Application class.

Similar proposal was in the FireDroid papaer~\cite{russello2013firedroid}. Their work was some kind of firewall between the applications and the system calls that works over the \textbf{ptrace()} system call. The further job is done by the monitor that attaches to apps and checks the policy statements.

Also there is a work that combines both statical and dynamic analysis methods~\cite{chen2013contextual}. It also acts as a monitor. But firstly it performs the static analysis and constructs Permission Event Graph, based on which all the work is done.

The other approach is an enforcement of the system rather than the applications. Some of those techniques were presented in~\cite{wang2016taming}. This work introduces SeApp application that performs an app instrumentation for tracking the inter-component communications.

 \chapter{Future Work}
\label{ChapFutureWork}

The possible future research for this topic covers many aspects. As long as this paper proposes the proof-of-concept code there is no entire listing in the app of potentially dangerous calls. This is not an easy task because although the Android project is open-sourced, a lot of API didn't documented~\cite{au2012pscout}

Beside that the additional feature for this method will be improving the speed of patching. Because the patching is based on merging 2 dex files it can take some time for this process. Therefore a ways for improving this method possibly exist.

Also the newer version of the DVM allow to use multidex format, so that the code could be placed in multiple \textit{classes\*.dex} files.

Because the most recent version of the Android prefers to use ART instead of DVM the continuations of the work is the performing patching of the native libraries.

\bibliographystyle{plain}
\bibliography{thesis}

\appendix
\chapter{Methods signatures of c}
\label{appendix:class_methods}

\begin{lstlisting}[language=Java]
// type pool of internal-represented strings
List<String> tnames = dexFile.typeNames();
// type indexes
List<Integer> types = dexFile.typeIds();
// strings table
List<String> strings = dexFile.strings();
// field indexes
List<FieldId> fields = dexFile.fieldIds();
// prototypes indexes
List<ProtoId> protos = dexFile.protoIds();
// method indexes
List<MethodId> methods = dexFile.methodIds();

for (Method m : c.methods()) {
    int i = m.getMethodIndex();
    MethodId id = methods.get(i);

    String n = strings.get(id.getNameIndex());
    ProtoId proto = protos.get(id.getProtoIndex());

    String ret = tnames.get(proto.getReturnTypeIndex());
    short types = dexFile.readTypeList(
        proto.getParametersOffset()).getTypes());

    print(ret + " " + n + " (");
    for (short t: types)
      print(tnames.get(t) + " ");
    print(") ");
}
\end{lstlisting}

\chapter{Stub geneneration based on dex}
\label{appendix:stub}

\begin{lstlisting}[language=Java]
public <T> void generateMethod(String name, Type<T> ret, Type...params) {
    Type<?> stub = Type.get("Lru/innopolis/Stub;");
    MethodId methodId = stub.getMethod(ret, name, params);
    Code code = generator.declare(methodId, 
        Modifier.STATIC | Modifier.PUBLIC);
    if (ret.equals(Type.VOID)) {
        code.returnVoid();
    } else {
        MethodId<T, Void> ctr = ret.getConstructor();
        Local<T> r = code.newLocal(ret);
        code.newInstance(r, ctr);
        code.returnValue(r);
    }
}
\end{lstlisting}

\end{document}